\documentclass[11pt,a4paper]{article}
\pdfoutput=1
\usepackage{jheppub}

\usepackage{bm}
\usepackage{graphicx}

\def\be{\begin{equation}}
\def\ee{\end{equation}}
\def\bear{\begin{eqnarray}}
\def\eear{\end{eqnarray}}
\def\nn{\nonumber}

\begin{document}


\title{Entanglement entropy on the fuzzy sphere}
\author{Joanna L. Karczmarek}
\author{and Philippe Sabella-Garnier}

\affiliation{Department of Physics and Astronomy, University of British Columbia, \\ 6224 Agricultural Road,
Vancouver, Canada}

\emailAdd{joanna@phas.ubc.ca}
\emailAdd{psabella@phas.ubc.ca}

\abstract{
We obtain entanglement entropy on the noncommutative (fuzzy) two-sphere.
To define a  subregion with a well defined boundary in this geometry, 
we use the symbol map between elements of the noncommutative algebra 
and functions on the sphere.
We find that entanglement entropy is not proportional
to the length of the region's boundary.  Rather, 
in agreement with holographic predictions,
it is extensive for regions whose area is a small (but fixed)
fraction of the total area of the sphere. This is true even in the limit
of small noncommutativity. We also find that entanglement
entropy grows linearly with $N$, where $N$ is the size of the irreducible 
representation of SU(2) used to define the fuzzy sphere.
}
\arxivnumber{1310.8345}
\maketitle

\section{Introduction}

Field theories on noncommutative spaces are interesting for
many reasons, one of them being that these theories
are inherently nonlocal and as such might serve as toy models
for certain phenomena in quantum gravity.  For example,
it has been conjectured that black hole horizons scramble information
so fast that in the membrane paradigm the horizon of a black hole 
cannot be modeled by a local theory \cite{Sekino:2008he}.
It has been suggested that noncommutative theories
might in fact be examples of fast scramblers, since (at least
at strong coupling) they exhibit enhanced thermalization rates compared
to local theories \cite{Edalati:2012jj}.

Noncommutative gauge theories arise naturally in string theory 
as the effective theories for low energy degrees of freedom
on a D-brane with a worldvolume magnetic field \cite{Seiberg:1999vs}.
Using this fact, it is possible
\cite{Hashimoto:1999ut,Maldacena:1999mh} to find
holographic duals to such theories.  Intriguingly, 
the dual geometry is an ordinary manifold (though not asymptotically
AdS), with noncommutativity  encoded in the shape of the 
holographic dual.  Since the dual has ordinary geometry,
it can be used to study intrinsically geometric observables
such as the geometric entanglement entropy
(entanglement entropy associated with
some region of space and obtained by tracing out all degrees of
freedom residing outside of this region).
If we are able to provide an interpretation for such 
holographically defined geometric observables
in the noncommutative field theory, we can study
how noncommutative space emerges both from the gravitational
dual and the noncommutative algebra on the field theory side.
Entanglement entropy is a particularly interesting 
geometric observable as it provides information on how degrees of 
freedom at different points are coupled, probing
nonlocality of the theory and perhaps teaching us
about its scrambling behaviour (for example, it was shown in
\cite{Lashkari:2013iga} that possessing extensive entanglement
entropy is a necessary condition for scrambling).  

Recent work \cite{Fischler:2013gsa,Karczmarek:2013xxa}
studied holographic entanglement entropy and mutual information,
uncovering some interesting properties of these observables in 
strongly coupled noncommutative gauge theories.\footnote{See also 
an earlier work \cite{Barbon:2008ut}.}
In particular, \cite{Karczmarek:2013xxa} argued
that UV/IR mixing leads to extensive (volume-law)
holographic entanglement entropy, instead of the more
usual area-law behaviour.  To interpret these findings within
field theory, we must answer the following questions:
Can one divide the Hilbert space of 
a field theory on some noncommutative geometry
into two components associated with the inside and the 
outside of some geometric region?
If not, what precisely is the meaning of holographic entanglement entropy
in field theory, and if yes, 
is  the volume-law behaviour observed through a holographic description
a property associated with strong coupling
or would it be also seen at weak coupling?
This last question is further motivated by the 
fact that, for example, the enhancement in  
thermalization timescale mentioned above  is not seen in perturbation theory
\cite{Edalati:2012jj}.

In the present paper, we shed some light on these 
issues by considering one of the
simplest nontrivial noncommutative field theories: the theory for a
free scalar on a noncommutative (or fuzzy) two-sphere.\footnote{The theory
of a free scalar on a noncommutative plane is equivalent
to the free scalar on a commutative plane and therefore
not interesting.}$^,$\footnote{Previous work on entanglement entropy
on a fuzzy sphere includes \cite{Dou:2006ni,Dou:2009cw}, where
the entanglement entropy for half the sphere was computed.}
The main advantage of working with the 
noncommutative sphere is that the field theory
is UV-finite and expressible as a finite size 
matrix model.  Further, the matrix model for a free scalar is
purely quadratic and therefore entanglement entropy is
straightforward to compute.

To obtain the theory of a real scalar on a noncommutative
two-sphere, we embed
the sphere in three dimensions and represent the three
Cartesian coordinates by SU(2) generators in the N dimensional
irreducible representation \cite{Madore:1992,Douglas:2001ba}:
\be
X^a = R \frac{L^a}{\sqrt{j(j+1)}}~,~~[L^a, L^b] =  i \epsilon^{abc} L^c~,
\ee
where the spin $j=(N-1)/2$.  Since
\be
L^a L^a = j(j+1) = \frac{N^2 - 1}{4}~,
\ee
the radius of the sphere is $R$.  A real scalar field $\varphi$ living on
the sphere is described by a hermitian $N\times N$ matrix $\Phi$.
Since $L^a$ are generators
of rotations, derivatives on the sphere are given by commutators
with $L^a$ and the Laplacian acting on a field $\varphi$ is
\be
-\frac{1}{R^{2}}~ [L^a, [L^a, \Phi]]~.
\ee
Since trace replaces integration over the sphere 
\be
 R^2 \int \sin(\theta) d\theta d\phi ~\varphi(\theta, \phi)
~~~ \rightarrow~~~
\frac{4\pi R^2}{N} ~\mathrm{Tr}~ \Phi~,
\ee
our free field has the following Hamiltonian
\begin{equation}
H=\frac{4 \pi R^2}{N}~
\frac{1}{2}~\mathrm{Tr}\left(\dot{\Phi}^2-R^{-2}[L_i,\Phi]^2+\mu^2\Phi^2\right)~.
\label{H}
\end{equation}
The theory has one free dimensionless parameter, mass measured in units of the 
radius, or $m = R \mu$.

In this paper, we obtain entanglement entropy associated with a polar
cap region $C$ whose size is controlled by a polar angle $\theta$
(see figure \ref{f1}).  
Using a map from operators (matrices) to functions on the sphere---called 
a symbol \cite{Douglas:2001ba}---we 
determine which entries of the matrix $\Phi$
correspond to degrees of freedom inside this polar cap and
which correspond to the outside.  Thus, we write the Hilbert space
$\cal H$ of our matrix model as a product of two smaller spaces: 
one corresponding to the inside of the polar cap, ${\cal H}_{C}$ 
and one corresponding to the outside of the polar cap,
${{\cal H}}_{\bar C}$, with  
${\cal H} = {\cal H}_{C} \otimes  {{\cal H}}_{\bar C}$.
We compute entanglement entropy using the usual definition:
\be
S = -\mathrm{Tr}_C ~\left (\rho_C \ln \rho_C\right )~,
\ee
where
\be \rho_c = \mathrm{Tr}_{\bar C}~ |\psi \rangle \langle \psi |
\ee
is the density matrix associated~with ${\cal H}_c$ when the 
entire quantum system is in a state $|\psi\rangle$ (which
we will take to be the vacuum).

Of course, it is not possible to draw a sharp 
boundary for a region on a noncommutative
sphere.  For our procedure, the boundary
of the region $C$ can be thought of as having a thickness
of $\sqrt{\bm\theta}$ where $\bm\theta = R^2/N$ is the 
noncommutativity parameter.  We can compare
the length-scale $\sqrt{\bm\theta} = R/\sqrt N$ to the
the UV cutoff of the theory, $\epsilon$.
The UV cutoff is most easily obtained  by dividing the area 
of the sphere $4\pi R^2$ by the total
number of degrees of freedom in our noncommutative model, $N^2$.
Since a small region of area $\epsilon^2$ should contain
exactly one degree of freedom, $\epsilon$ is approximately $R/N$. 
We see that $\sqrt{\bm\theta}$ is parametrically larger than
$\epsilon$.

Since, in the large $N$ limit, the noncommutative sphere 
is supposed to reduce to the commutative one,
the reader might expect entanglement entropy
on a noncommutative sphere to agree with that on
a commutative sphere for regions whose diameter is 
larger than $\sqrt{\bm\theta}$.   If that were the case,
we would not be able to discover any deviation in the noncommutative 
case, as regions whose boundary has thickness $\sqrt{ \bm \theta}$ cannot
be smaller than $\sqrt{\bm\theta}$.  
Fortunately, as
has been observed in \cite{Fischler:2013gsa,Karczmarek:2013xxa},
strong deviations from commutative behaviour should
be seen in entanglement entropy for regions whose size
is of order  $\bm \theta / \epsilon = R$, which
corresponds to the entire sphere (or, equivalently, 
the IR cutoff of our theory).  
The reason why noncommutative entanglement entropy for 
regions larger than $\sqrt{\bm \theta}$  (but smaller than
$\bm \theta / \epsilon$)
does not agree with its commutative counterpart
lies in UV/IR mixing: because of this mixing,
a noncommutative theory with a UV cutoff $\epsilon$ is
expected to have nonlocal behaviour up to a length-scale
$\bm\theta/\epsilon$ \cite{Minwalla:1999px}\footnote{
In \cite{Minwalla:1999px}, the UV/IR connection was studied
on the noncommutative plane.  The UV/IR mixing on the fuzzy
sphere has been studied using the one-loop effective action
in several interacting theories (see for example \cite{Chu:2001xi} and 
\cite{CastroVillarreal:2004vh}).  
Here we simply use the results from flat noncommutative
geometry as a guide to interpreting our results.
Notice that entanglement entropy could potentially be sensitive to
UV/IR mixing not detected by, for example, divergences in the 
two-point functions.  In this case our results could be 
interpreted as evidence of previously undiscovered UV/IR mixing.}.  
Thus, we expect deviation
from commutative behaviour at least for regions whose 
area is a small (but finite in the commutative limit)
fraction of the total sphere area.
Any such deviation we see can be interpreted as
a result of UV/IR mixing on the noncommutative sphere.

In fact, this is precisely what we discover:
for small regions, entanglement entropy on the noncommutative sphere
grows linearly with the area of the region (and not with
the length of its boundary), and hence follows the volume law.\footnote{
Even though we are working in two spacial dimensions, we will
continue to use higher-dimensional terminology and
refer to entropy growing with the area of the region as
volume-law behaviour and  entanglement
entropy proportional to the length of the boundary of the region
as area-law behaviour.}  For regions whose area
is comparable to the total area of the sphere, the
entanglement entropy receives higher power corrections.
However, while in \cite{Karczmarek:2013xxa} it was
shown that the entanglement entropy for a field theory with
some effective noncommutativity scale $a_\theta$ at strong coupling
undergoes a phase transition between volume-law at length-scales
below $a_\theta^2 / \epsilon$ and area-law at length-scales above
that, on the noncommutative sphere there is no such phase 
transition.  This is due to the compactness of the manifold and the
resulting IR cutoff which was absent in the holographic 
calculation.  The phase transition is replaced with 
crossover behaviour near the IR cutoff (for a region whose
size is half that of the whole sphere) and the higher power corrections
mentioned above lead to a smooth behaviour.

The rest of this paper is organized as follows: In section \ref{ncg}
we present a self-contained review of symbol maps 
and star products on the noncommutative plane
and the noncommutative sphere, before presenting our
proposal for how to define the polar cap region in the noncommutative sphere 
geometry in section \ref{polar-cap}.  
Our numerical methodology and results follow in section \ref{results}.
Finally, in section \ref{discusion} we provide possible
interpretations for our results and discuss open problems for future work.

\section{Noncommutative geometry}
\label{ncg}

In this section, we review the concepts of a symbol and a corresponding
star product as a way to encode the  noncommutative structure of geometry.
We begin by reviewing the better-known example of a noncommutative plane, and then show how the 
same tools can be applied to treat the noncommutative sphere.
Our general approach is similar to that in
\cite{Alexanian:2000uz}, though the details are somewhat different.
In section \ref{polar-cap} we use our symbol map on the fuzzy sphere
to obtain our desired mapping between the polar cap and matrix elements.

\subsection{Noncommutative plane}

The noncommutative plane has as its structure algebra the
Heisenberg algebra.  This is the algebra generated by two
operators $\hat x$ and $\hat y$ with the commutation
relation
\be
[\hat x, \hat y] = i {\bm \theta}~.
\ee
$\sqrt{\bm \theta}$ has the units of length and is the fundamental
length-scale of noncommutativity.  

A common treatment of noncommutative geometry 
uses a map $s$ which takes elements $\hat A$ of the structure algebra
to functions on the corresponding commutative manifold
(in this case, the ordinary two-dimensional plane), 
$s(\hat A) =  f_A(x,y)$.
The function $f_A$ is called the symbol of the algebra element $\hat A$.
The symbol map is not unique: there are many different definitions
of $s$, corresponding to different ways to order the algebra element
$s^{-1}(f_A)$.  
For every symbol map $s$ there exists a so-called star product $*$
with the property that
\be
s(\hat A \hat B) = s(\hat A) * s(\hat B)~.
\ee
As $\hat A \hat B \neq \hat B \hat A$, $*$ cannot be a commutative product,
but it is associative.  

The most often used symbol map is the Weyl-ordered symbol, which leads
to the Moyal star product.  Explicitly, these are
\be
s_W(\hat A)  = \frac{1}{(2\pi)^2} \int d^2k ~\mathrm{Tr}
~\left (e^{ik\cdot(x-\hat x)} \hat A\right )
\ee
and
\be
(f \star g)(x,y) ~= e^{\frac{i}{2}{\bm \theta} \left(
\frac{\partial}{\partial \xi_1}
\frac{\partial}{\partial  \zeta_2} - 
\frac{\partial}{\partial \zeta_1}
\frac{\partial}{\partial  \xi_2} \right )} ~ 
f(x+\xi_1,y+\zeta_1) g(x+\xi_2,y+\zeta_2)~
|_{\xi_1 = \zeta_1 = \xi_2 = \zeta_2 = 0}~.
\label{star-prod}
\ee
However, this is not the treatment we wish to present.  Instead,
we follow the approach due to Berezin \cite{Berezin:1974du} and define the
symbol as an expectation value in a coherent state.

Accordingly, let us define the raising and lowering operators, 
$\hat a = \hat x + i\hat y$, $\hat a^\dagger = \hat x - i\hat y$,
with $[\hat a, \hat a^\dagger] = 2{\bm\theta}$, as well as the canonical 
coherent states which are eigenstates of the lowering
operator, $\hat a|\alpha\rangle = \alpha|\alpha\rangle$.
We can think of each coherent state $|\alpha\rangle$ as corresponding
to the point $(x,y)=(\mathrm{Re}(\alpha),\mathrm{Im}(\alpha))$ since 
$\langle \alpha | \hat x| \alpha \rangle = \mathrm{Re}(\alpha)$ and 
$\langle \alpha | \hat y| \alpha \rangle = \mathrm{Im}(\alpha)$.
The overlap between these coherent states decreases
rapidly as the corresponding points are separated:
$|\langle \beta |\alpha\rangle| =  e^{-|\alpha-\beta|^2/4{\bm\theta}}$,
with a length-scale controlled by the noncommutativity scale $\sqrt {\bm\theta}$.
The Berezin symbol is defined as
\be
s(\hat A) = f_A(\alpha, \bar \alpha) =  \langle \alpha | \hat A | \alpha \rangle~.
\ee
The Berezin symbol corresponds to normal ordering,
in contrast with the Weyl symbol which was based on the symmetric
ordering. 

To derive a star product compatible with the Berezin symbol,
we need to use the fact that $|\alpha\rangle$ is, up to a 
normalization factor, a holomorphic function of the complex 
variable $\alpha$:
\be
|\alpha \rangle = e^{-|\alpha|^2/4 {\bm\theta}}~\sum_{n=0}^\infty \frac{(\alpha
/\sqrt{2{\bm\theta}})^n}
{\sqrt{n!}} |n \rangle~.
\ee
This implies that $\langle \beta | A | \alpha \rangle 
/\langle \beta | \alpha \rangle $ is holomorphic in $\alpha$ and
antiholomorphic in $\beta$.  Thus, 
since ${\langle \beta |  \beta \rangle } = 1$, we have
\be
\frac {\langle \beta | A | \alpha \rangle }
{\langle \beta |  \alpha \rangle } = 
e^{-\beta \frac{\partial}{\partial \alpha} }
\frac {\langle \beta | A | \beta +\alpha \rangle }
{\langle \beta |  \beta + \alpha \rangle } = 
e^{-\beta \frac{\partial}{\partial \alpha} }
e^{\alpha \frac{\partial}{\partial \beta} }
\frac {\langle \beta | A | \beta  \rangle }
{\langle \beta |  \beta \rangle } = 
e^{-\beta \frac{\partial}{\partial \alpha} }
e^{\alpha \frac{\partial}{\partial \beta} } 
f_A(\beta, \bar \beta)~,
\label{holomorphic}
\ee
and, similarly,
\be
\frac {\langle \alpha | A | \beta \rangle }
{\langle \alpha |  \beta \rangle } = 
e^{-\bar \beta \frac{\partial}{\partial \bar \alpha} }
e^{\bar \alpha \frac{\partial}{\partial \bar \beta} } 
f_A(\beta, \bar \beta)~.
\ee

We can now compute an explicit expression for the star product,
using the completeness relation for coherent states:
\bear
(f_A * f_B) (\beta, \bar \beta) &=& 
\langle \beta |  A B| \beta \rangle = 
\frac{1}{2\pi{\bm\theta}}\int d^2\alpha  
\langle \beta | A| \alpha \rangle
\langle \alpha |  B |\beta \rangle
\nn \\ &=&  
\frac{1}{2\pi{\bm\theta}}\int d^2\alpha~
|\langle \beta |  \alpha \rangle|^2
~
\left [ e^{-\beta \frac{\partial}{\partial \alpha} }
e^{\alpha \frac{\partial}{\partial \beta} } 
f_A(\beta, \bar \beta) \right ]~
\left [e^{-\bar \beta \frac{\partial}{\partial \bar \alpha} }
e^{\bar \alpha \frac{\partial}{\partial \bar \beta} } 
f_B(\beta, \bar \beta) \right ]~
\nn \\ &=&  
\frac{1}{2\pi{\bm\theta}}\int d^2\alpha~
\left (e^{\beta \frac{\partial}{\partial \alpha} +
\bar \beta \frac{\partial}{\partial \bar \alpha} }
|\langle \beta |  \alpha \rangle|^2 \right )
~
\left [ 
e^{\alpha \frac{\partial}{\partial \beta} } 
f_A(\beta, \bar \beta) \right ]~
\left [
e^{\bar \alpha \frac{\partial}{\partial \bar \beta} } 
f_B(\beta, \bar \beta) \right ]
\nn \\ &=&  
\frac{1}{2\pi{\bm\theta}}\int d^2\alpha~
|\langle \beta |  \alpha +\beta\rangle|^2 
~
\left [ 
e^{\alpha \frac{\partial}{\partial \beta} } 
f_A(\beta, \bar \beta) \right ]~
\left [
e^{\bar \alpha \frac{\partial}{\partial \bar \beta} } 
f_B(\beta, \bar \beta) \right ]
\nn \\ &=&  
\frac{1}{2\pi{\bm\theta}}\int d^2\alpha~
e^{-|\alpha|^2/2{\bm\theta}}
~
\left [ 
e^{\alpha \frac{\partial}{\partial \beta} } 
f_A(\beta, \bar \beta) \right ]~
\left [
e^{\bar \alpha \frac{\partial}{\partial \bar \beta} } 
f_B(\beta, \bar \beta) \right ]
\nn \\ &=&  
e^{2{\bm\theta} \frac{\partial}{\partial \zeta} \frac{\partial}{\partial \bar \eta} } 
f_A(\beta+\zeta, \bar \beta) 
f_B(\beta, \bar \beta+\bar\eta)  |_{\zeta=\bar \eta = 0}
\eear
This is known as the Vorol product.  The Vorol product is 
equivalent to the more commonly used Moyal star product shown above.

\subsection{Noncommutative sphere}

A rather similar approach allows us to study the noncommutative
sphere.  The structure algebra is simply the algebra
of $N\times N$ hermitian matrices, $M_n$.  In this algebra, 
as we have already discussed in the Introduction,
we single out three matrices $L^a$ satisfying the SU(2) commutation relations.
Since the $L^a$ form an irreducible representation of SU(2),
these three matrices generate all of $M_n$.

We will use as a basis the eigenvectors of the $L^3$ 
angular momentum, $|m\rangle$:
\be
L^3 |m\rangle = m|m\rangle~,~~m=-j\ldots j~,~~ \langle  m | m \rangle =1~,
~~j=\frac{N-1}{2}~.
\ee

To define an analog of the coherent state,\footnote{For another approach to coherent states on the fuzzy sphere, see \cite{Hammou:2002ky}} let $\hat n$ be a unit 3-vector 
(or a point on a unit sphere).  Then, define $L_{\hat n} = \hat n^a L^a$.
A coherent state at point $R\hat n$ on the sphere of radius $R$
is then $|\hat n \rangle$, where
\be
L_{\hat n} |\hat n \rangle = j |\hat n \rangle~,~~
 \langle  \hat n | \hat n \rangle =1~.
\ee
The coherent state at the north pole is the state with the largest
angular momentum in the 3-direction; a coherent state at any 
other point can be obtained from the one at the north pole 
by a SU(2) rotation.  Recall the Wigner formula,\footnote{
See for example \cite{Sakurai:1994book} equation (3.8.33).}
\be
|\langle m | \hat n \rangle| = \sqrt {\frac{(2j)!}{(j+m)!)(j-m)!}} ~
\left ( \cos  \frac{\theta}{2} \right )^{j+m} ~ 
\left ( \sin  \frac{\theta}{2} \right )^{j-m}~,
\label{wigner}
\ee
where $\theta$ is the polar angle at point $\hat n$ 
on the unit sphere: the
angle between the positive $3$-axis and 
$\hat n$.

One consequence is that, if the angle between two
unit vectors $\hat n_1$ and $\hat n_2$ is $\chi =
\arccos ( \hat n_1 \cdot \hat n_2)$, then
\be
|\langle \hat n_1 | \hat n_2 \rangle| = 
\left ( \cos  \frac{\chi}{2} \right )^{2j}  = 
\left ( \frac{\hat n_1 \cdot \hat n_2+1}{2} \right )^j~. 
\ee
For large $j$, the overlap between the states 
$|\hat n_1 \rangle$ and $|\hat n_2 \rangle$ 
decreases sharply as the angle between them is increased.
If we let $\chi = 2 / \sqrt j$, we have
\be
\left (  \cos  \frac{\chi}{2} \right )^{2j}  \approx
\left (  1 - \frac{1}{2j} \right )^{2j}  ~,
\ee
which approaches $1/e$ for large $j$.  Thus,
the effective width of the coherent states on a sphere
of radius R is proportional
to $RN^{-1/2}$.  A single coherent state covers an area proportional
to $R^2/N$, which is natural given that the noncommutative
sphere should contain $N$ unit noncommutative `cells'.\footnote
{In string theory, we would say that the spherical
D2-brane has $N$ units of flux piercing it, corresponding to
$N$ D0-branes dissolved in its worldvolume.  The effective
theory describing $N$ D0-branes is written in terms of $N\times N$
hermitian matrices, lending the D2-brane the noncommutative
structure we are studying.}

It is easy to convince oneself that for $\hat n$ in the 1-3 plane,
$\langle m | \hat n \rangle$ can be 
real when we take $L_2$ to be purely imaginary (and therefore
antisymmetric).  To restore the phase of 
$\langle m | \hat n \rangle$ for all directions $\hat n$, 
write $\hat n$ in polar coordinates:
$\hat n = (\sin \theta \cos \phi, \sin\theta\sin\phi,\cos\theta)$.
As we just discussed, for the azimuthal angle $\phi=0$, $\langle m | \hat n \rangle$ is real.  
For all other angles, we rotate around the 3-axis to obtain
\be
\langle m | \hat n \rangle = \sqrt {\frac{(2j)!}{(j+m)!)(j-m)!}} ~
\left (\frac {1}{2}\sin \theta \right )^{j} ~ 
\left ( \tan  \frac{\theta}{2} \right )^{-m} e^{-im\phi} ~.
\label{wigner2}
\ee
Now, consider a complex variable $\alpha = R \tan \left ( \theta/2 \right )
e^{i\phi}$. This is simply the complex coordinate arising from
a stereoscopic projection.  This coordinate does not
cover the entire sphere (it is singular at the point $\theta=\pi$),
but a complementary complex coordinate, 
$\tilde \alpha = R \tan \left ( (\pi-\theta)/2 \right ) e^{-i\phi}$,
does.  Since $\tilde \alpha(\alpha) = R^2/\alpha$ is a 
holomorphic function,  together these two complex
coordinates define a complete complex structure.

We will now change notation and denote the coherent 
states with $|\alpha\rangle$ instead of $|\hat n\rangle$.
Just as it was with the coherent state on the plane,
up to a normalization factor our coherent states on the sphere
are holomorphic in the  complex variable $\alpha$.
\be
| \alpha \rangle  = \left (\frac {1}{2}\sin \theta \right )^{j} ~ 
\sum_{m=-j}^{j}~
\sqrt {\frac{(2j)!}{(j+m)!)(j-m)!}} 
~~ \left (\frac{\alpha}{R}\right )^{-m} |m\rangle~.
\ee
These coherent states are overcomplete
\be
\frac{N}{4 \pi R^2}\int \frac {4 d^2\alpha}{(1+|\alpha/R|^2)^2}
~| \alpha \rangle \langle \alpha |  ~=~ 1 ~,
\ee
with respect to the  SU(2) invariant measure on the sphere,
$\frac {4d^2\alpha}{(1+|\alpha/R|^2)^2}$.

For any matrix operator $A$ in $M_n$, consider its Berezin symbol 
$f_A(\alpha) = \langle \alpha | A | \alpha  \rangle$.
The Berezin symbol is a function on the sphere which
corresponds to the matrix in $M_n$.  Since equation (\ref{holomorphic})
is valid on the sphere (as it relies only on the
coherent states being holomorphic), we  have:
\bear
(f_A * f_B) (\beta, \bar \beta) &=& 
\langle \beta |  A B| \beta \rangle ~=~ 
\frac{N}{\pi R^2}\int \frac {d^2\alpha}{(1+|\alpha/R|^2)^2}
\langle \beta | A| \alpha \rangle
\langle \alpha |  B |\beta \rangle
 \\ \nn &=&  
\frac{N}{\pi R^2}\int \frac {d^2\alpha}{(1+|\alpha/R|^2)^2}
|\langle \beta |  \alpha \rangle|^2
~
\left [ e^{-\beta \frac{\partial}{\partial \alpha} }
e^{\alpha \frac{\partial}{\partial \beta} } 
f_A(\beta, \bar \beta) \right ]~
\left [e^{-\bar \beta \frac{\partial}{\partial \bar \alpha} }
e^{\bar \alpha \frac{\partial}{\partial \bar \beta} } 
f_B(\beta, \bar \beta) \right ]~.
\eear

To simplify our computation, we will
compute  only the star product at the north pole, $\beta=0$.
We have
\bear
(f_A * f_B) (0,0) &=& 
\frac{N}{\pi R^2}\int \frac {d^2\alpha}{(1+|\alpha/R|^2)^2}~
|\langle 0 |  \alpha \rangle|^2
~
\left [ 
e^{\alpha \frac{\partial}{\partial \beta} } 
f_A(\beta, \bar \beta) \right ]_{\beta=0}~
\left [
e^{\bar \alpha \frac{\partial}{\partial \bar \beta} } 
f_B(\beta, \bar \beta) \right ]_{\beta=0}~
 \\ &=&   \nn
\frac{N}{\pi R^2}\int \frac {d^2\alpha}{(1+|\alpha/R|^2)^2}~
\left ( \frac{1}{1+|\alpha/R|^2}\right )^{2j}
~
\left [ 
e^{\alpha \frac{\partial}{\partial \beta} } 
f_A(\beta, \bar \beta) \right ]_{\beta=0}~
\left [
e^{\bar \alpha \frac{\partial}{\partial \bar \beta} } 
f_B(\beta, \bar \beta) \right ]_{\beta=0}~.
\eear
Now, on the surface of it, this integral does not 
appear convergent; however, we have:
\be
\frac{\partial^p}{\partial \beta^p} f(\beta,\bar \beta) |_{\beta=0}=
0~~~~\mathrm{for~}{p > 2j}~.
\ee
To see that this is the case, just write the Berezin symbol for
any operator $A$ as
\bear
f_A(\beta,\bar \beta) &=& \sum_{n,m=-j}^j \langle \beta | n \rangle
\langle n | A |m \rangle \langle m | \beta \rangle 
\nn \\ &=&
\sum_{n,m=-j}^j \frac {(2j)! ~~\langle n | A |m \rangle}
{{\sqrt{(j+m)!(j+n)!(j-n)!(j-m)!}}} 
\frac{\beta^{j-m} \bar \beta^{j-n}}{(1+\beta \bar \beta)^{2j}}
\eear
 $\frac{\partial^p}{\partial \beta^p}$ acting on a $(n,m)$ term
in the above sum is nonzero only if $p=j-m$ and $n=j$.  Thus,
$p$ is at most $2j$.

Returning to our expression for the star product, we now have
that
\bear
(f_A * f_B) (0,0) &=& \frac{N}{\pi R^2} \sum_{p,q=0}^{2j} \frac{1}{p!q!}
\int \frac {d^2\alpha ~\alpha^p \bar \alpha^q}
{\left (1+|\alpha/R|^2\right )^{2j+2}}~
\left [ \frac{\partial^p}{\partial \beta^p}
f_A(\beta, \bar \beta) \right ]_{\beta=0}~
\left [ \frac{\partial^q}{\partial \bar \beta^q}
f_B(\beta, \bar \beta) \right ]_{\beta=0}
\nn \\ &=&  
N~ \sum_{p=0}^{2j} R^{2p} \frac{(2j-p)!}{p!(2j+1)!}
\left [ \frac{\partial^p}{\partial \beta^p}
f_A(\beta, \bar \beta) \right ]_{\beta=0}~
\left [ \frac{\partial^p}{\partial \bar \beta^p}
f_B(\beta, \bar \beta) \right ]_{\beta=0}
\nn \\ &=&  
\sum_{p=0}^{2j}  \frac{(2j-p)!}{p!(2j)!}
\left [R^{2p} \frac{\partial^p}{\partial \alpha^p}
\frac{\partial^p}{\partial \bar \beta^p}
f_A(\alpha, \bar \alpha) 
f_B(\beta, \bar \beta) \right ]_{\alpha=\beta=0}
\eear
This is the star product derived and used, for example, 
in \cite{Presnajder:1999ky}.

If the functions $f_A$ and $f_B$ are very smooth, only the first
few terms will contribute.  We can then write
\bear
(f_A * f_B) (0,0) &\approx & 
 \sum_{p=0} \frac{(2j)^{-p}}{p!}
\left [ R^{2p}\frac{\partial^p}{\partial \alpha^p}
\frac{\partial^p}{\partial \bar \beta^p}
f_A(\alpha, \bar \alpha) 
f_B(\beta, \bar \beta) \right ]_{\alpha=\beta=0} 
\nn \\ &=& 
e^{\frac{R^2}{2j} \frac{\partial}{\partial \alpha}
\frac{\partial}{\partial \bar \beta}}
f_A(\alpha, \bar \alpha) 
f_B(\beta, \bar \beta) |_{\alpha=\beta=0} ~,
\eear
which reduces to the Vorov product with the 
noncommutativity parameter $R^2/(4j)$.

\begin{figure}
\center{
\includegraphics[scale=0.9]{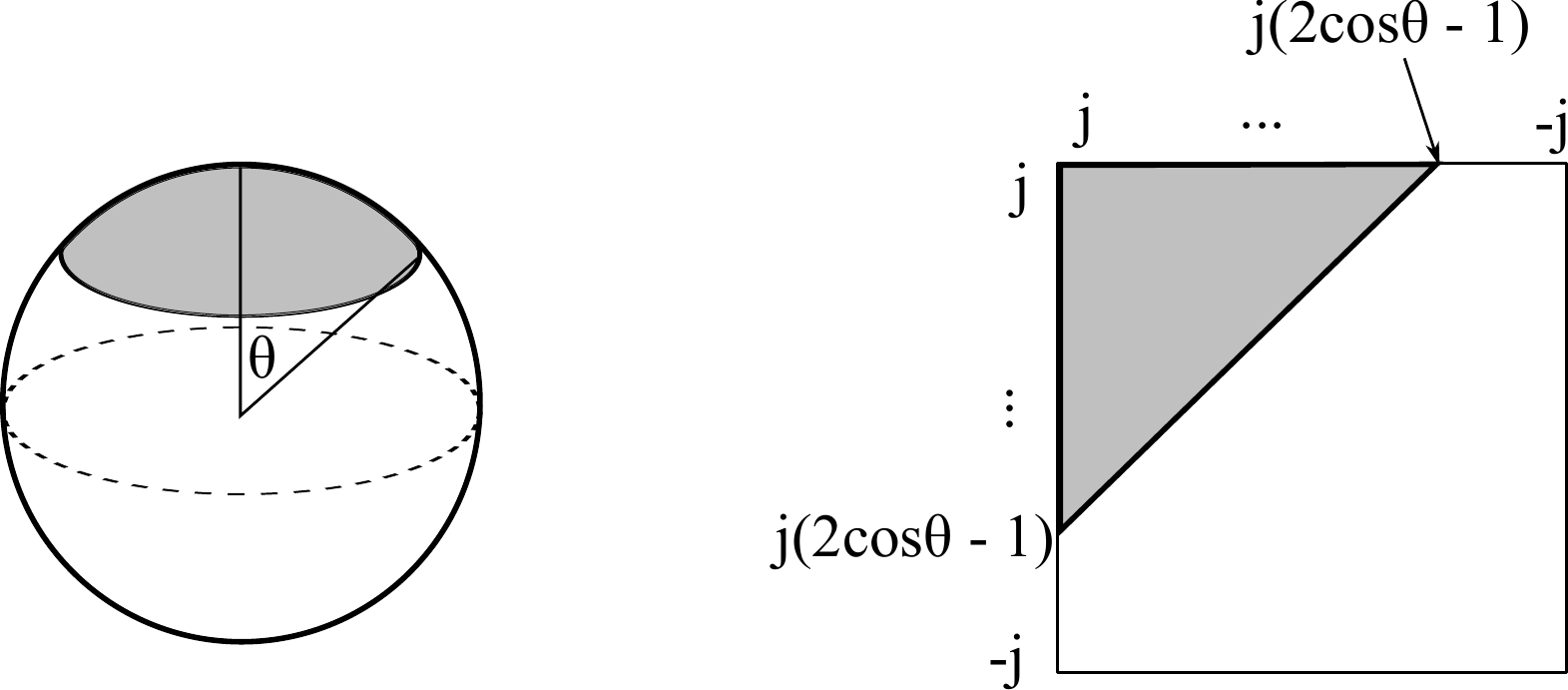}}
\caption{Degrees of freedom on the sphere and their matrix
counterparts.}
\label{f1}
\end{figure}

\subsection{The polar cap on the noncommutative sphere}
\label{polar-cap}

Now, consider an operator $|m_1\rangle\langle m_2|$
and its Berezin symbol $f_{m_1,m_2}(\alpha) =  
\langle \alpha |m_1\rangle\langle m_2| \alpha  \rangle$.
We have
\be
f_{m_1,m_2} =  (\textrm{phase})~
 {\frac{(2j)!}{\sqrt{(j+m_1)!)(j-m_1)!(j+m_2)!)(j-m_2)!}}} ~
\left ( \cos  \frac{\theta}{2} \right )^{2j+(m_1+m_2)} ~ 
\left ( \sin  \frac{\theta}{2} \right )^{2j-(m_1+m_2)}
\label{symbol-nm}
\ee
For large $j$, the function
\be
\left ( \cos  \frac{\theta}{2} \right )^{2j(1+x)} ~ 
\left ( \sin  \frac{\theta}{2} \right )^{2j(1-x)}~.
\ee
has a sharp peak at $\theta$ such that $\cos \theta = x$.  
Therefore the Berezin symbol $f_{m_1,m_2}(\alpha)$  is largest
when the vector $\hat n$ makes an angle 
$\theta_0 =\arccos{\left (\frac{m_1+m_2}{2j}\right)}$ with the 
vertical axis.
For $\theta$ close to $\theta_0$, we can write
\be
\left ( \cos  \frac{\theta}{2} \right )^{2j(1+x)} ~ 
\left ( \sin  \frac{\theta}{2} \right )^{2j(1-x)}~\approx~
\left [ \frac{1}{4} \left ( 1+x\right )^{1+x} 
\left ( 1-x\right )^{1-x}\right ]^j  e^{-j(\theta - \theta_0)^2}~.
\ee
Therefore, the Berezin symbol of the operator
$|m_1\rangle\langle m_2|$ is appreciable only when the polar
angle $\theta$ is within $1/\sqrt{j}$ of $\theta_0$.

This implies that the degrees of freedom corresponding
to a polar cap $C$ of angular radius $\theta$ (i.e., all points
on the sphere whose polar angle is less than  $\theta$) can be 
identified, in the large $j$ limit, with the set of matrix elements
$\lbrace \langle m_1 | \Phi | m_2 \rangle 
~|~ m_1+m_2 > 2j \cos(\theta) \rbrace$.  In particular,
to compute the entanglement entropy for half the sphere,
we should include the degrees of freedom in `half' the matrix.
This was conjectured, but not proven, in \cite{Dou:2006ni}.
Note that it does not matter whether the (anti)diagonal degrees of
freedom are included or not, as the answer will be the same in
any pure state.

Since our coherent states have a width proportional to $R /\sqrt N$,
the boundary of our polar cap region $C$ can be thought as having a thickness
of the same size, $R/\sqrt N$.  
In other words, if we consider the subspace of $M_n$ spanned just by the 
matrix elements indicated in figure \ref{f1}, the corresponding
functions on the sphere would have support on the polar cap $C$ and
Gaussian drop-off  `tails' controlled by $R/\sqrt N$
outside of the polar cap $C$.

Since the full set of symbols given by equation (\ref{symbol-nm})
have wavelengths as short as $R/j$ (natural, given that
the UV cutoff $\epsilon = R/N$), one can wonder whether it is
possible to fine-tune our procedure to produce a region with
a `thinner' boundary.  We leave this for future work.

\section{Results}
\label{results}

Since our Hamiltonian (\ref{H}) is quadratic, we use
the formalism developed in \cite{Srednicki:1993im} to
numerically compute entanglement entropy.  

We label the entries of $\Phi$ as\footnote{
We have chosen the precise computational method we are about to
describe for its conceptual simplicity.  The parametrization developed in 
\cite{Dou:2006ni} is a more efficient (numerically) approach for this problem.
The two approaches gave the same answers.}
\begin{equation}
\Phi=\left( \begin{array}{ccccc}
\Phi_1 & \frac{\Phi_2+i\Phi_3}{\sqrt{2}} & \frac{\Phi_4+i\Phi_5}{\sqrt{2}} & \frac{\Phi_7+i\Phi_8}{\sqrt{2}} & \dots  \\
\frac{\Phi_2-i\Phi_3}{\sqrt{2}} & \Phi_6 & \frac{\Phi_9+i\Phi_{10}}{\sqrt{2}} & \dots & \dots \\
\frac{\Phi_4-i\Phi_5}{\sqrt{2}} & \frac{\Phi_9-i\Phi_{10}}{\sqrt{2}}  & \dots & \dots & \dots \\
\frac{\Phi_7-i\Phi_8}{\sqrt{2}} & \dots  & \dots & \dots & \dots \\
\dots & \dots & \dots & \dots & \dots 
\end{array} \right) 
\end{equation}
so that the Hamiltonian (\ref{H}) takes the form 
\begin{equation}
H=\frac{2\pi}{N}\sum_{a,b=1}^{N^2} \left( \pi_a \delta_{ab} \pi_b + \Phi_a K_{ab} \Phi_b \right)~,
\end{equation}
where $K$ is the real symmetric positive-definite dynamical matrix 
\begin{equation}
K_{ab}=-\frac{1}{2}\frac{\partial^2 \mathrm{Tr}([L_i,\Phi]^2)}{\partial \Phi_a \partial \Phi_b}+m^2 \delta_{ab}
\end{equation}
and $\pi_a=R \dot{\Phi}_a$ are canonical momenta.

Note that if $m=0$, $K$ has a zero eigenvalue associated with the 
matrix $\Phi$ being proportional to the identity.  This flat direction leads
to infinite entanglement entropy.
To study the massless case, one could
impose a tracelessness condition on $\Phi$, thus eliminating the massless mode. 

\begin{figure}
\center{
\includegraphics[scale=0.3]{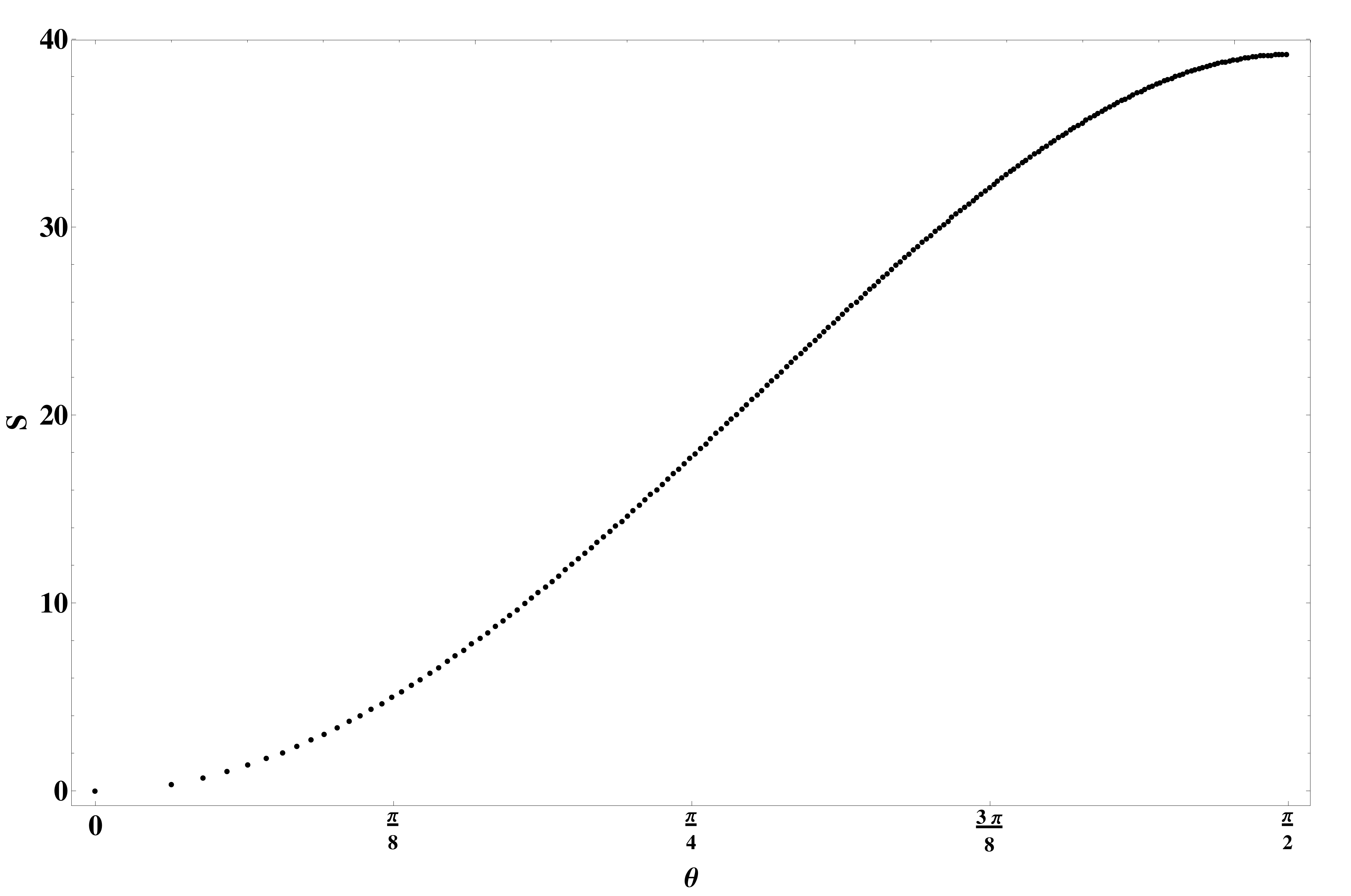}}
\caption{Entanglement entropy $S$ as a function of 
angular size $\theta$ of polar cap $C$. $N=200$ and $m=1$.}
\label{f-200}
\end{figure}

To calculate the entanglement entropy associated with a polar cap we must
compute the entanglement entropy associated with tracing over Hilbert spaces
of oscillators $\Phi_a$ in the first $k$ (anti)diagonal `lines', with
$k = 2j(1-\cos\theta)$.  Therefore, we need to find the entanglement entropy
of the first $n = 1+2+3+\ldots+k$ oscillators ($a=1,\ldots,n$), which
we can accomplish by following the procedure outlined in \cite{Srednicki:1993im}.
Since there is a bit of ambiguity in the relationship between $\theta$ and k 
($\theta$ can be shifted by an amount order $1/N$), we use this  to
adjust the correspondence between $\theta$ and $k$ so that 
the physical condition $S(\theta) = S(\pi-\theta)$ is satisfied.  
This is possible because entanglement entropies for $n=k(k+1)/2$ and $n=N^2-k(k+1)/2$
are the same (a good check on our numerics).  Thus
\begin{equation}
\cos \theta=1-\frac{k}{N-\frac{1}{2}}~,~~\mathrm{for}~ k=1, \dots, N-1.\\
\end{equation}

Figure \ref{f-200} shows entanglement entropy for a polar cap region 
as a function of the polar angle $\theta$ for $m=1$ and $N=200$.  Angles
beyond $\theta=\pi/2$ are not shown as entanglement entropy necessarily has
$S(\theta) = S(\pi-\theta)$ for a pure state such as the vacuum.  
The most interesting feature is the small angle behaviour: $S \sim \theta^2$.
Notice also that $S(\theta)$ is smooth as a function of $\theta$,
including at $\theta=\pi/2$, indicating
that there is no phase transition in the entanglement entropy.
Such a phase transition was observed for ${\cal N} = 4$ Yang-Mills
in \cite{Fischler:2013gsa,Karczmarek:2013xxa}, but here it is absent,
probably due to the theory being defined on a compact manifold.

\begin{figure}\center{
\includegraphics[scale=0.3]{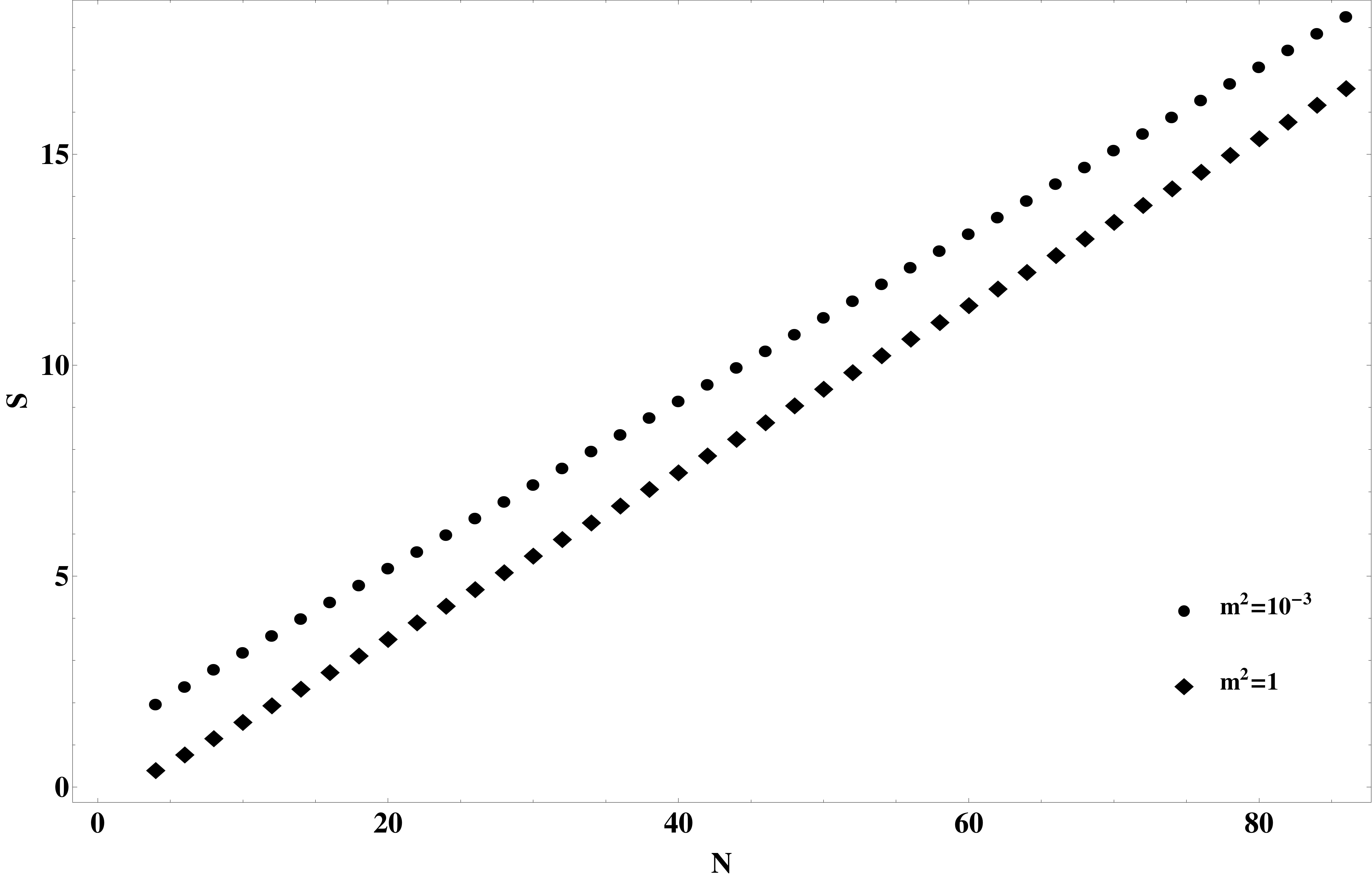}}
\caption{Entanglement entropy for half the sphere as a function of N for $m^2=1$
and $m^2=0.001$. Both lines have a slope of 0.2.}
\label{f-linear}
\end{figure}

In figure \ref{f-linear} we show the dependence of the
entropy $S$ for a half-sphere ($\theta=\pi/2$) as a function of $N$.
The behaviour is clearly linear (though with an offset dependent
on the mass), leading us to conclude that
\be
\frac{S}{N} =  F(\theta) ~+~ {\cal O}\left({N^{-1}}\right )\mathrm{corrections}~
\label{result}
\ee
where $F(\theta)$ is proportional to $\theta^2$ for small $\theta$, is smooth for $\theta \in [0,\pi]$ and has the property that $F(\pi-\theta)=F(\theta)$. 
\begin{figure}\center{
\includegraphics[scale=0.3]{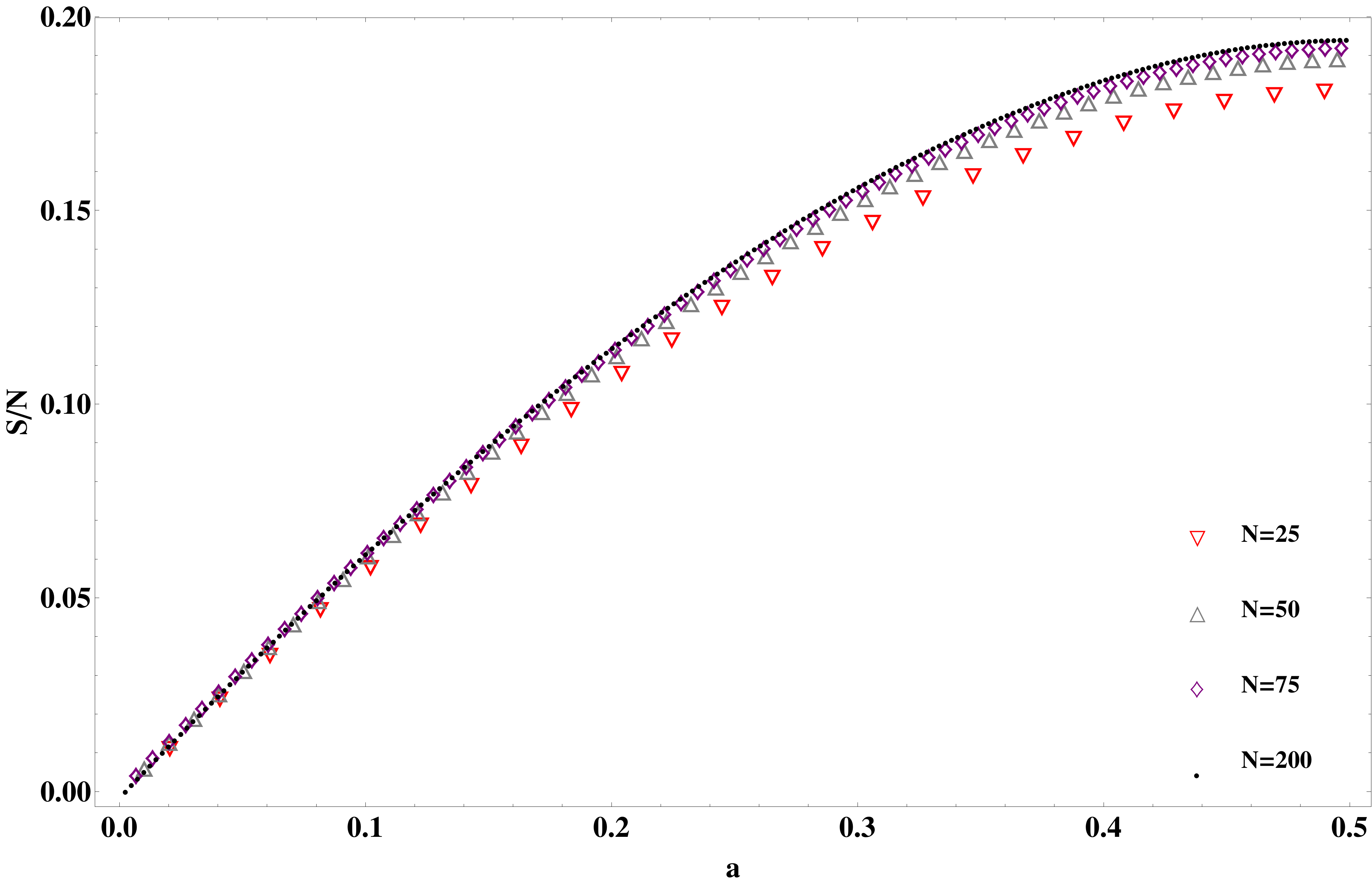}}
\caption{Entanglement entropy $S$ scaled by $N$ as the fractional area $a$ of polar cap for different
values of $N$.  Notice that $S/N$ converges to a good large $N$ limit.
For small $\theta$ $S/N$ appears proportional to $C$, consistent with an extensive entanglement entropy. 
 $m=1$.}
\label{f-largeN}
\end{figure}

The claim in equation (\ref{result}) is further supported by figure \ref{f-largeN}, which shows
$S/N$ as a function of the fractional area of the polar cap,
\be
a:=\frac{\mathrm{area}(C)}{4\pi R^2}= \sin^2(\theta/2)~,
\ee
 for several different values of $N$.  The convergence
to a fixed curve at large $N$ is evident.

Result (\ref{result}) is quite reasonable from the point
of view of the matrix model.  We can think of the square matrix
in figure \ref{f1} as literally a square block of coupled harmonic oscillators.
In this way of thinking, the couplings arising from Hamiltonian (\ref{H})
are only among nearest-neighbour oscillators.  Therefore, we would
expect the entanglement entropy to follow an area law.  The length of the 
boundary is the number of matrix elements laying on the diagonal 
in figure \ref{f1}, which is $2j(1-\cos\theta) = 2 N a$, thus we
expect the entanglement entropy in this square array of 
oscillators to be proportional to $N a$, which is what we see
for small $a$ in figure $\ref{f-largeN}$.

\begin{figure}\begin{center}
\parbox{5.0in}{
\includegraphics[scale=0.3]{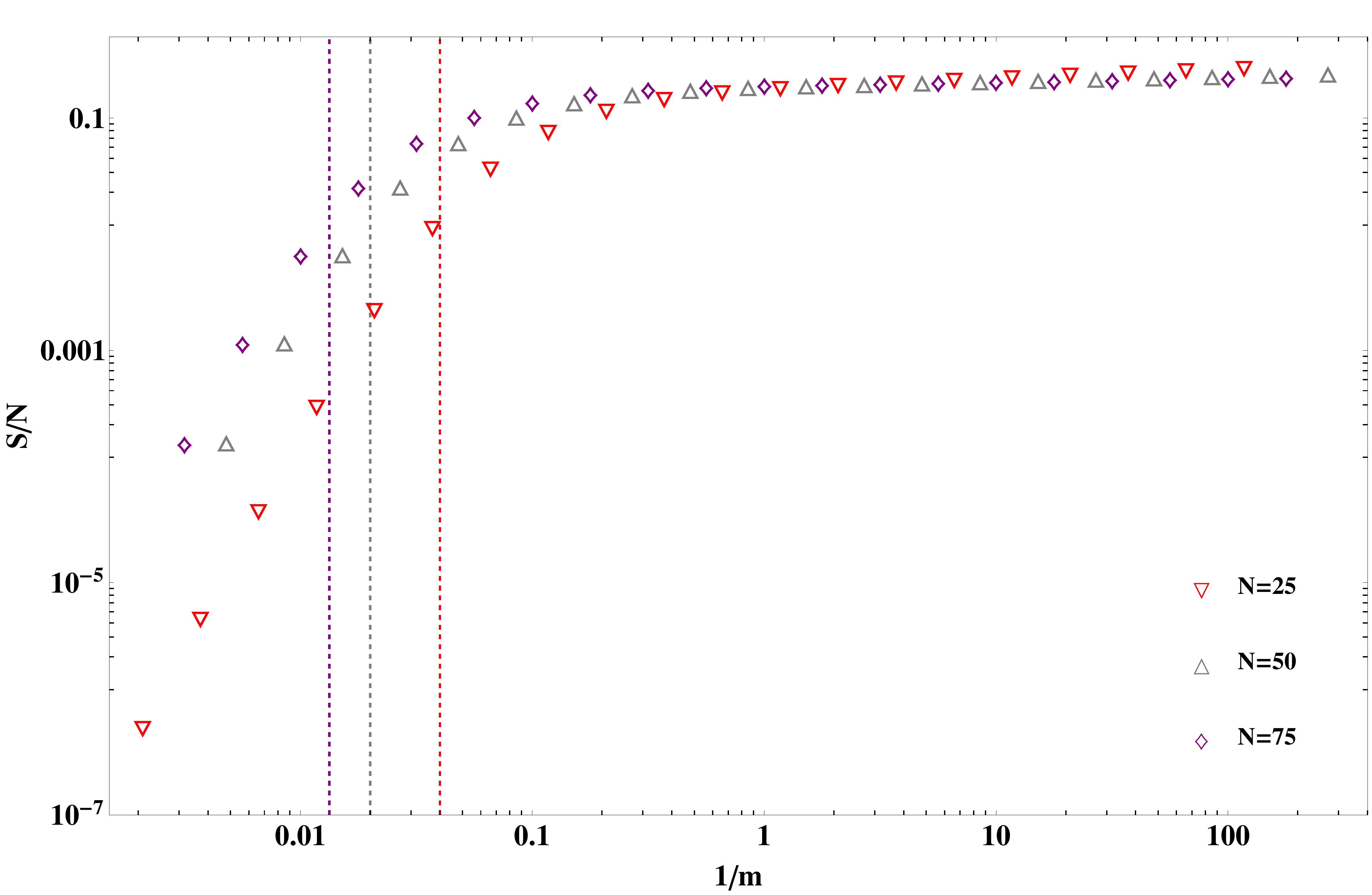} 
\vskip -2.1in
\hskip 2.0in
\includegraphics[scale=0.12]{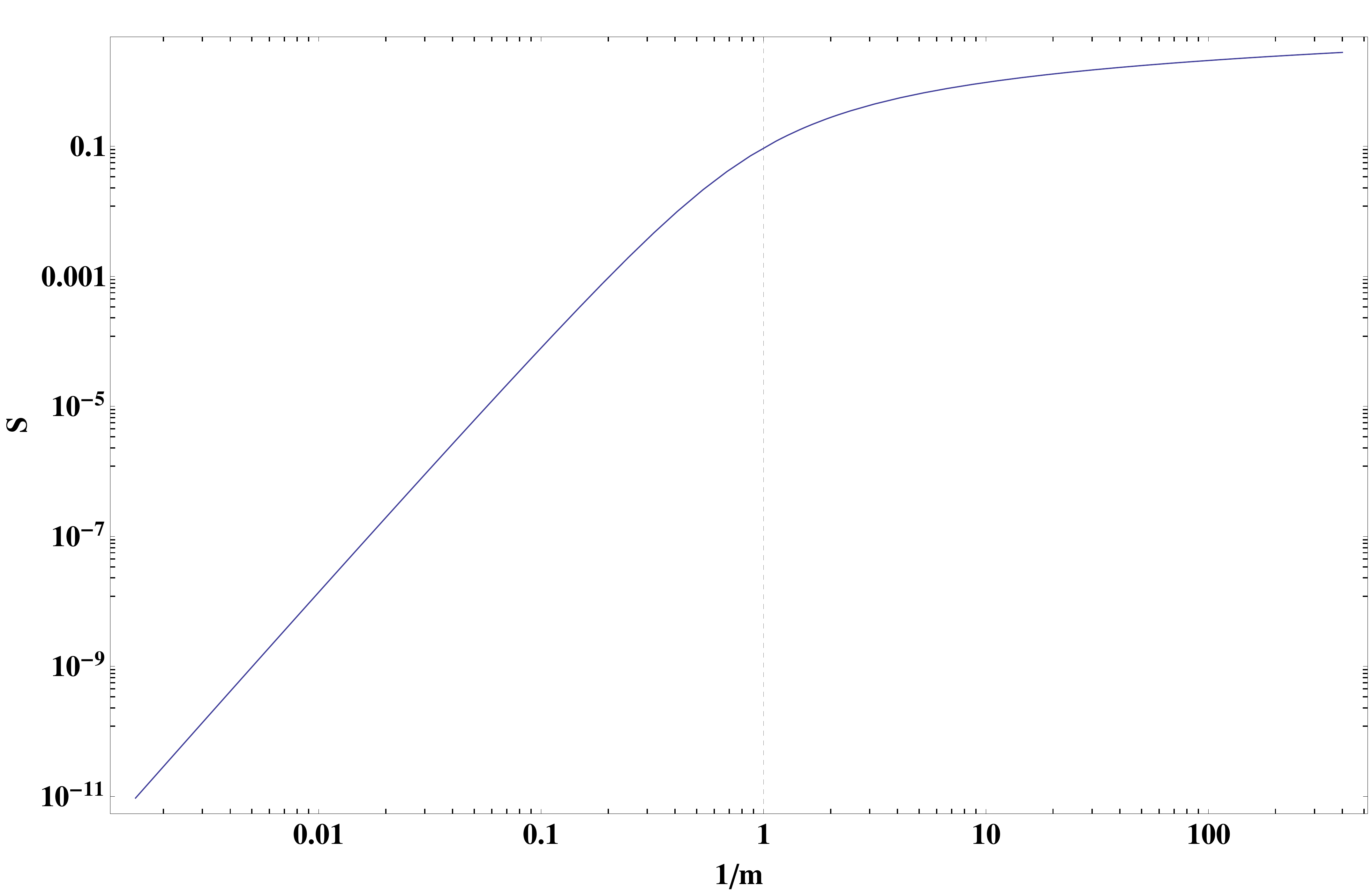}}\end{center}
\vskip 0.5in
\caption{Entanglement entropy for half the sphere as a function of inverse mass.
Vertical lines indicate  mass at which the diagonal and off-diagonal 
elements of the Hamiltonian (\ref{H}) matrix are of the same order ($m \sim N$).
The inset shows entanglement entropy for two coupled harmonic oscillators whose potential
energy is given by equation (\ref{Htoy}).}
\label{f-mass}
\end{figure}

Finally, we studied dependence of the entanglement entropy on the dimensionless mass $m$ of the field.  
Figure \ref{f-mass} shows the entanglement entropy for half the sphere as a function of
$m^{-1}$.  Over a wide range of masses, for $m<N$, the entanglement entropy
appears approximately independent of the mass.  This is the region in which
our result (\ref{result}) is applicable.  In fact, in this region, $S$ must slowly raise
with $m^{-1}$, since at $m=0$ the entropy is infinite due to the appearance of a 
flat direction in the model.  For large masses $m>N$, the entropy decreases to zero,
as the kinetic term (which couples degrees of freedom at different points and is
the source of entanglement entropy) is overwhelmed by the mass term
(which does not couple degrees of freedom at different points).
To understand this behaviour, let's consider a toy model of two coupled oscillators
$x_1$ and $x_2$ (of equal masses) coupled by a potential 
\be
(1+m^2) x_1^2 + (1+m^2) x_2^2 + 2 x_1 x_2~.
\label{Htoy}
\ee
Entanglement entropy for one of these oscillators, in the vacuum of the system and
as a function of $m^{-1}$ is shown in the inset to figure \ref{f-mass}.  This toy example
models the full Hamiltonian (\ref{H}): the kinetic term in (\ref{H}) contributes
diagonal and off-diagonal terms (of order 1 in the toy Hamiltonian (\ref{Htoy})),
while the mass term contributes only diagonal terms.  We see that the behaviour
of the entanglement entropy in this toy model has the same qualitative features as
the entanglement entropy on the sphere. 

\begin{figure}\center{
\includegraphics[scale=0.3]{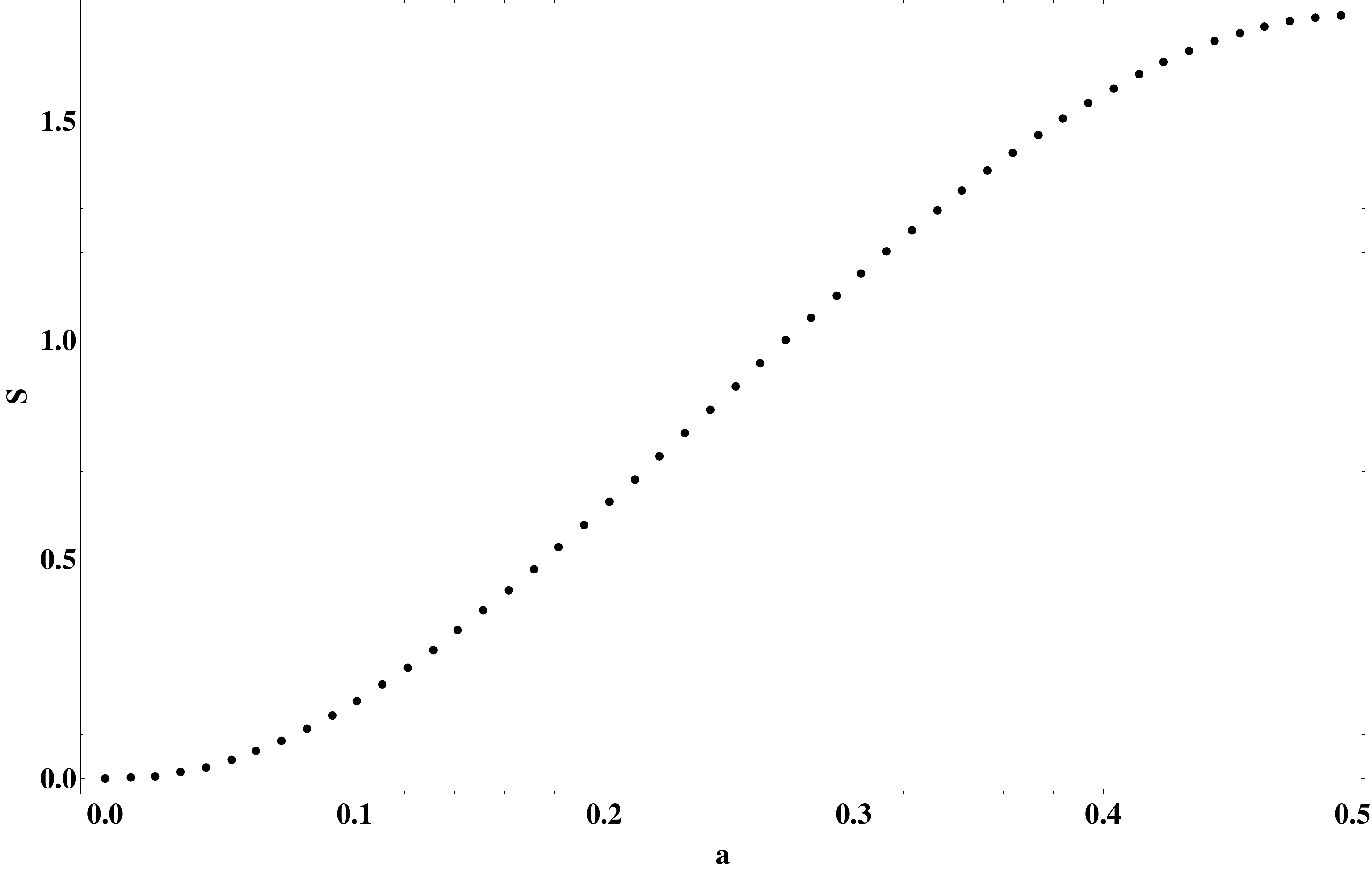}}
\caption{Entanglement entropy $S$ scaled by $N$ as the fractional area $a$ of polar cap 
for a large mass, $m^2=1000$ ($N$=50).}
\label{f-large-m}
\end{figure}

When we study entanglement entropy for a  field with a large mass, we find that it
grows with $N$ faster than linear, and that its behaviour as a function of $\theta$ 
at small $\theta$ and at fixed $N$ is $\theta^a$, with a power $a>2$ 
(figure \ref{f-large-m}).  Notice that in
the limit of infinite mass, the harmonic oscillators that make up our noncommutative
sphere appear effectively uncoupled, so it is not surprising that entanglement
 entropy is smaller at a large mass.  We leave an exploration of the details of
this behaviour to future work.

\section{Discussion and future work}
\label{discusion}

An interesting point about our construction of a region in
noncommutative geometry, which is illustrated in figure \ref{f1}, is
that the number of oscillators in the triangular corner of
the matrix is approximately $2 j^2 (1-\cos\theta)^2$ or
$2 N^2 a^2$.  Thus, this number of degrees of freedom
associated with a polar cap does not grow proportionately 
to its fractional area $a$, but rather to the square of $a$.
This would not be possible in a local theory (where degrees of
freedom are more-or-less associated with points in the underlying
geometry).  It also implies that there is no reason why the
degrees of freedom associated with a subregion of the sphere
should be uniformly distributed within this subregion.
If degrees of freedom in a subregion of a noncommutative
sphere reside predominantly near the boundary of the region,
we can account for our entanglement entropy
results as follows: 
just like in commutative theories, quantum correlations between 
a region and its complement develop only across the boundary,
but since these degrees of freedom are concentrated near the boundary,
entanglement entropy  grows more rapidly than it would in a commutative theory.
Such a `reshuffling' of degrees of freedom
when a finite region is considered would be an interesting
way in which a noncommutative field theory can reproduce
the entanglement entropy predicted by holography.

\begin{figure}\center{
\includegraphics[scale=0.3]{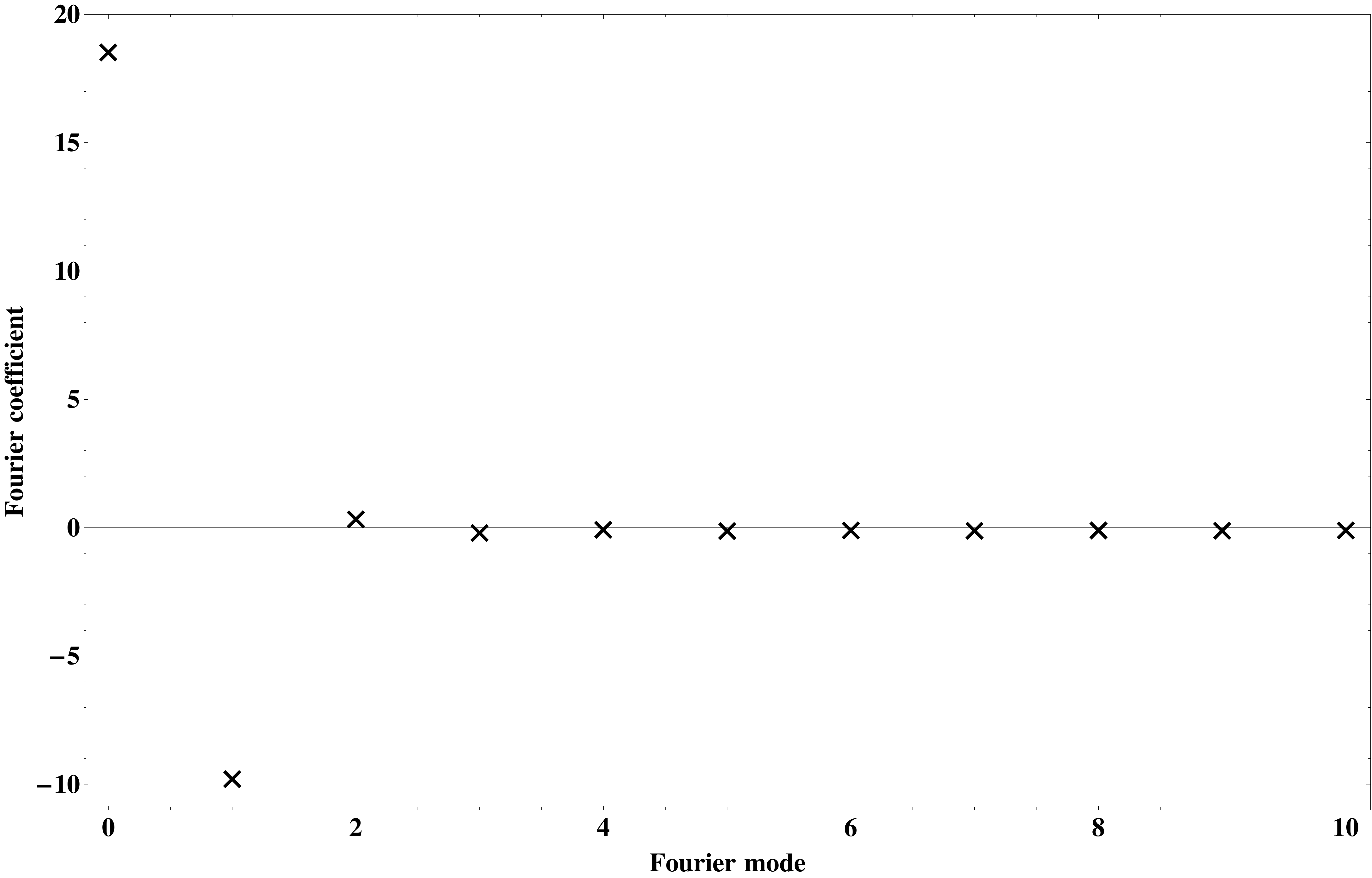}}
\caption{Coefficients of $\cos(2n\theta)$ terms in a Fourier 
expansion for entanglement entropy $S$ shown in figure \ref{f-200}. Since entanglement entropy is symmetric about $\theta=\frac{\pi}{2}$, the Fourier coefficients $c_n$ satisfy $c_n=c_{400-n}$ for $n=1\dots 199$, which implies $S(\theta)=c_0+2\sum_{n=1}^{199}c_n \cos(2n\theta)+c_{200}\cos(2\cdot 200 \cdot \theta)$.}
\label{f-fourier}
\end{figure}

An alternative interpretation of our results can be obtained
when we realize that
our result for the entanglement entropy can be written in
some suggestive ways.  In the linear regime, we have that
\be
S =  \kappa N ~\frac{\mathrm{area}(C)}{R^2}  = 
\kappa ~ \frac{\mathrm{area}(C)}{\bm\theta} ~.
\label{S1}
\ee
This should be compared with the general form of the  result from 
\cite{Karczmarek:2013xxa}, which was
\be
S \sim   \frac{\mathrm{area}(C)}{\epsilon^2} ~.
\label{S2}
\ee
It is as if the UV cutoff $\epsilon$ was replaced by $\sqrt {\bm \theta}$.
Such a replacement would be fairly natural in the case of area-law entropy, as $\sqrt {\bm \theta}$
is the thickness of the boundary;  however, it is hard to see how the
fuzziness of a boundary definition can affect entropy  in the 
volume-law regime.  It would be very interesting to
see what happens to the entanglement entropy if one can
define a region with a boundary whose thickness is less than $\sqrt {\bm \theta}$,
ideally as small as $\epsilon$.

A further intriguing fact about the entanglement entropy we have found is that
$F(\theta)$ in equation (\ref{result})
is very close to a multiple of $\sin^2(\theta)$.
This is presented in figure \ref{f-fourier}; the
first three coefficients are 18.6, -9.68 and 0.443. The 
ratio of the first two coefficients is -1.92 $\approx$ -2,
which leads to $F(\theta) \sim \sin^2(\theta) = a(1-a)$,
where $a=\sin^2(\theta/2)$ is the fractional area of the polar cap.
The expression 
\be
S \approx N\kappa a (1-a)~,~~\kappa = 0.8
\label{S3}
\ee
is most suggestive.  If degrees of freedom are pictured
as uniformly distributed on the sphere, the dependence of $S$ on
the area $a$ is most easily explained by assuming that every (localized) degree
of freedom on the sphere has its entanglement uniformly spread over
the whole sphere.  This is in sharp contrast with local
theories, where short-ranged interactions lead to
quantum correlations between a region and its complement
being established across the boundary surface, leading to
the area-law for entanglement entropy \cite{Eisert:2008ur}.
In this interpretation, however,
it is puzzling why the
entanglement entropy does not grow like the total number of degrees of freedom, $N^2$.
Perhaps the answer is that the localized degrees of freedom are only weakly
entangled with each other.  Or perhaps the $N$-dependence is
caused by the thickness of the boundary (notice that replacing
$N$ with $N^2$ in equation (\ref{S3}) is equivalent to changing
$\bm \theta$ to $\epsilon^2$ in equation (\ref{S1}), which turns it
into equation (\ref{S2})).

To distinguish between these different interpretations of our
results, further study is needed.  In addition to 
finding a procedure for building regions with thinner boundaries,
one could also study mutual information.  Since mutual information
is UV finite in local theories, it might have a different dependence
on the UV and other cutoffs in the noncommutative theory.

Our method for obtaining finite cap region on the noncommutative
sphere can also be applied to disk-like regions in the noncommutative plane.
It would be interesting to study entanglement entropy in that geometry,
(necessarily) for an interacting field theory, and to see 
whether entanglement entropy in a field theory really does have 
a phase transition at a length-scale ${\bm\theta}/\epsilon$, in
accordance with holographic results. Moreover, the theory on a noncommutative plane
is related to the quantum Hall effect \cite{Bigatti:1999iz}.
Other geometries that would be interesting to study would be 
higher dimensional spheres as well as  noncommutative
tori in different dimensions.


\section*{Acknowledgments}

We are grateful for helpful discussions with Charles
Rabideau, Gordon Semenoff,  Mark van Raamsdonk and Ken Yeh.  This work was completed 
with support from the Natural Sciences and Engineering Council
of Canada (NSERC) and  from the
Fonds de recherche du Qu\'{e}bec --- Nature et technologies
(FRQNT).

\bibliographystyle{JHEP}
\bibliography{new}

\providecommand{\href}[2]{#2}\begingroup\raggedright\begin{thebibliography}{10}

\bibitem{Sekino:2008he}
Y.~Sekino and L.~Susskind, {\it {Fast Scramblers}},  {\em JHEP} {\bf 0810}
  (2008) 065, [\href{http://xxx.lanl.gov/abs/0808.2096}{{\tt
  arXiv:0808.2096}}].

\bibitem{Edalati:2012jj}
M.~Edalati, W.~Fischler, J.~F. Pedraza, and W.~Tangarife~Garcia, {\it {Fast
  Scramblers and Non-commutative Gauge Theories}},  {\em JHEP} {\bf 1207}
  (2012) 043, [\href{http://xxx.lanl.gov/abs/1204.5748}{{\tt
  arXiv:1204.5748}}].

\bibitem{Seiberg:1999vs}
N.~Seiberg and E.~Witten, {\it {String theory and noncommutative geometry}},
  {\em JHEP} {\bf 9909} (1999) 032,
  [\href{http://xxx.lanl.gov/abs/hep-th/9908142}{{\tt hep-th/9908142}}].

\bibitem{Hashimoto:1999ut}
A.~Hashimoto and N.~Itzhaki, {\it {Noncommutative Yang-Mills and the AdS / CFT
  correspondence}},  {\em Phys.Lett.} {\bf B465} (1999) 142--147,
  [\href{http://xxx.lanl.gov/abs/hep-th/9907166}{{\tt hep-th/9907166}}].

\bibitem{Maldacena:1999mh}
J.~M. Maldacena and J.~G. Russo, {\it {Large N limit of noncommutative gauge
  theories}},  {\em JHEP} {\bf 9909} (1999) 025,
  [\href{http://xxx.lanl.gov/abs/hep-th/9908134}{{\tt hep-th/9908134}}].

\bibitem{Lashkari:2013iga}
N.~Lashkari, {\it {Equilibration of Small and Large Subsystems in Field
  Theories and Matrix Models}},  \href{http://xxx.lanl.gov/abs/1304.6416}{{\tt
  arXiv:1304.6416}}.

\bibitem{Fischler:2013gsa}
W.~Fischler, A.~Kundu, and S.~Kundu, {\it {Holographic Entanglement in a
  Noncommutative Gauge Theory}},  \href{http://xxx.lanl.gov/abs/1307.2932}{{\tt
  arXiv:1307.2932}}.

\bibitem{Karczmarek:2013xxa}
J.~L. Karczmarek and C.~Rabideau, {\it {Holographic entanglement entropy in
  nonlocal theories}},  {\em JHEP} {\bf 1310} (2013) 078,
  [\href{http://xxx.lanl.gov/abs/1307.3517}{{\tt arXiv:1307.3517}}].

\bibitem{Barbon:2008ut}
J.~L. Barbon and C.~A. Fuertes, {\it {Holographic entanglement entropy probes
  (non)locality}},  {\em JHEP} {\bf 0804} (2008) 096,
  [\href{http://xxx.lanl.gov/abs/0803.1928}{{\tt arXiv:0803.1928}}].

\bibitem{Dou:2006ni}
D.~Dou and B.~Ydri, {\it {Entanglement entropy on fuzzy spaces}},  {\em
  Phys.Rev.} {\bf D74} (2006) 044014,
  [\href{http://xxx.lanl.gov/abs/gr-qc/0605003}{{\tt gr-qc/0605003}}].

\bibitem{Dou:2009cw}
D.~Dou, {\it {Comments on the Entanglement Entropy on Fuzzy Spaces}},  {\em
  Mod.Phys.Lett.} {\bf A24} (2009) 2467--2480,
  [\href{http://xxx.lanl.gov/abs/0903.3731}{{\tt arXiv:0903.3731}}].

\bibitem{Madore:1992}
J.~Madore, {\it {The fuzzy sphere}},  {\em Class. Quantum Grav.} {\bf 9} (1992)
  69--87.

\bibitem{Douglas:2001ba}
M.~R. Douglas and N.~A. Nekrasov, {\it {Noncommutative field theory}},  {\em
  Rev.Mod.Phys.} {\bf 73} (2001) 977--1029,
  [\href{http://xxx.lanl.gov/abs/hep-th/0106048}{{\tt hep-th/0106048}}].

\bibitem{Minwalla:1999px}
S.~Minwalla, M.~Van~Raamsdonk, and N.~Seiberg, {\it {Noncommutative
  perturbative dynamics}},  {\em JHEP} {\bf 0002} (2000) 020,
  [\href{http://xxx.lanl.gov/abs/hep-th/9912072}{{\tt hep-th/9912072}}].

\bibitem{Chu:2001xi}
C.-S. Chu, J.~Madore, and H.~Steinacker, {\it {Scaling limits of the fuzzy
  sphere at one loop}},  {\em JHEP} {\bf 0108} (2001) 038,
  [\href{http://xxx.lanl.gov/abs/hep-th/0106205}{{\tt hep-th/0106205}}].

\bibitem{CastroVillarreal:2004vh}
P.~Castro-Villarreal, R.~Delgadillo-Blando, and B.~Ydri, {\it {A
  Gauge-invariant UV-IR mixing and the corresponding phase transition for U(1)
  fields on the fuzzy sphere}},  {\em Nucl.Phys.} {\bf B704} (2005) 111--153,
  [\href{http://xxx.lanl.gov/abs/hep-th/0405201}{{\tt hep-th/0405201}}].

\bibitem{Alexanian:2000uz}
G.~Alexanian, A.~Pinzul, and A.~Stern, {\it {Generalized coherent state
  approach to star products and applications to the fuzzy sphere}},  {\em
  Nucl.Phys.} {\bf B600} (2001) 531--547,
  [\href{http://xxx.lanl.gov/abs/hep-th/0010187}{{\tt hep-th/0010187}}].

\bibitem{Berezin:1974du}
F.~Berezin, {\it {General Concept of Quantization}},  {\em Commun.Math.Phys.}
  {\bf 40} (1975) 153--174.

\bibitem{Hammou:2002ky}
A.~B. Hammou, M.~Lagraa, and M.~Sheikh-Jabbari, {\it {Coherent State Induced
  Star-Product on $R^3_\lambda$ and the Fuzzy Sphere}},  {\em Phys. Rev.} {\bf
  D66} (2002) 025025, [\href{http://xxx.lanl.gov/abs/hep-th/0110291}{{\tt
  hep-th/0110291}}].

\bibitem{Sakurai:1994book}
J.~J. Sakurai and J.~Napolitano, {\it {Modern quantum mechanics}}, .

\bibitem{Presnajder:1999ky}
P.~Presnajder, {\it {The Origin of chiral anomaly and the noncommutative
  geometry}},  {\em J.Math.Phys.} {\bf 41} (2000) 2789--2804,
  [\href{http://xxx.lanl.gov/abs/hep-th/9912050}{{\tt hep-th/9912050}}].

\bibitem{Srednicki:1993im}
M.~Srednicki, {\it {Entropy and area}},  {\em Phys.Rev.Lett.} {\bf 71} (1993)
  666--669, [\href{http://xxx.lanl.gov/abs/hep-th/9303048}{{\tt
  hep-th/9303048}}].

\bibitem{Eisert:2008ur}
J.~Eisert, M.~Cramer, and M.~Plenio, {\it {Area laws for the entanglement
  entropy - a review}},  {\em Rev.Mod.Phys.} {\bf 82} (2010) 277--306,
  [\href{http://xxx.lanl.gov/abs/0808.3773}{{\tt arXiv:0808.3773}}].

\bibitem{Bigatti:1999iz}
D.~Bigatti and L.~Susskind, {\it {Magnetic fields, branes and noncommutative
  geometry}},  {\em Phys.Rev.} {\bf D62} (2000) 066004,
  [\href{http://xxx.lanl.gov/abs/hep-th/9908056}{{\tt hep-th/9908056}}].

\end{thebibliography}\endgroup

\end{document}